\documentclass[twocolumn,groupedaddress,aps,prl,floatfix,showpacs,a4]{revtex4}
\usepackage{amssymb}
\usepackage{amsmath}
\usepackage{graphicx}
\usepackage{color}

\begin{document}
\addtolength{\abovedisplayskip}{-1.5pt}
\addtolength{\columnsep}{-1pt}
\addtolength{\belowdisplayskip}{-2.5pt}
\addtolength{\textfloatsep}{-20pt}
\addtolength{\abovecaptionskip}{-10pt}
\addtolength{\parskip}{-0.5pt}

\title{Langevin equation with colored noise for \\constant-temperature molecular dynamics simulations}
\author{Michele Ceriotti}
\author{Giovanni Bussi}
\email{gbussi@unimore.it}
\altaffiliation{Present address: S3 research center
   and Dipartimento di Fisica, Universit\`a di Modena e Reggio Emilia, via Campi 213/A, 41100 Modena, Italy.
}
\author{Michele Parrinello}
\affiliation{Computational Science, Department of Chemistry and Applied Biosciences,
ETH Z\"urich, USI Campus, Via Giuseppe Buffi 13, CH-6900 Lugano, Switzerland}
\date{\today}

\begin{abstract}
We discuss the use of a Langevin equation with a colored (correlated) noise
to perform constant-temperature molecular dynamics simulations.
Since the equations of motion are linear in nature,
it is easy to predict the response of a Hamiltonian system to such a thermostat
and to tune at will the relaxation time of modes of different frequency. 
This allows one to optimize the time needed to thermalize the system and 
generate independent configurations. We show how this frequency-dependent response
can be exploited to control the temperature of Car-Parrinello-like 
dynamics, keeping at low temperature the electronic degrees of freedom, 
without affecting the adiabatic separation from the vibrations of the ions.
\end{abstract}

\vspace{-0.5cm}
\maketitle
Solving Hamilton's equations leads to sampling of the microcanonical 
constant-energy distribution, but in real-life experiments it is the temperature
that is kept constant. Reproducing this condition in computer
simulations is of great importance for the investigation of a large class of 
physical, chemical and biological problems.
Several approaches have been proposed to modify Hamilton's equations 
in order to perform constant-temperature 
dynamics (see e.g.~Refs.~\cite{THERMO,NOSE,CHAINS,buss+07jcp}).
Many of these~\cite{THERMO,buss+07jcp} rely on stochastic methods, which are a natural
choice for modeling the interactions with an external heat bath,
and which display excellent ergodic behavior due to their
random nature. A good thermostat should be able to rapidly enforce
the correct probability distribution, and generate 
uncorrelated configurations, which are necessary to compute ensemble averages.
The efficiency of the thermostat is particularly important in 
{\em ab initio} simulations, because of their high computational cost.
The stochastic thermostats used so far are based on Markovian equations of motion, 
and imply no memory of the past trajectory of the system.

Markovian random processes are, however, only a subset of all possible stochastic
processes. Furthermore, the Mori-Zwanzig theory ensures that whenever some 
degree of freedom is integrated out, the dynamics of the remaining degrees of 
freedom are described by a non-Markovian Langevin equation, with a finite-range
memory function~\cite{zwan61pr,mori65ptp,zwan+01book}. Hence, in the quest for a better thermostat, 
and considering the thermostat as arising from a set of bath variables whose 
effect is integrated out, it is natural to explore the effect of using a 
non-Markovian Langevin equation to perform constant-temperature molecular dynamics. 
In this Letter we will show that, by using colored noise,
it is possible to influence in a different manner the different vibrational modes 
of the system. Therefore the thermostat can be adjusted to the system under study,
and its performance optimized in a precise and predictable fashion.
This is, to our knowledge, the first time that a colored Langevin equation 
has been employed in atomistic simulations.

An area which would greatly benefit from an improved, tunable thermostat is that
of Car-Parrinello (CP)-like, extended Lagrangian schemes~\cite{car-parr85prl}.
The idea behind this approach is very general, as 
it applies to any system where the forces are the result of an 
expensive optimization procedure.
This process is circumvented by extending the dynamical 
degrees of freedom (DOF), so as to include the parameters to 
be optimized, and introducing an artificial dynamics which allows 
these extra variables to be maintained close to the ground state, by adiabatic
decoupling from the other degrees of freedom.
In the prototypical example of CP molecular dynamics (CPMD)
a fictitious mass is assigned to the electronic DOF
so that they can be evolved at the same time as the ionic
DOF. If the fictitious mass is small enough, the dynamics of the electrons
are adiabatically separated from the dynamics of the ions. Hence,
the electrons are kept close to the ground state, while the nuclei
are evolved at the correct temperature.
This same technique can be used in classical simulations that use polarizable
force fields, where the electronic DOF describe the charge polarization
of the system~\cite{spri91jcp,rick+94jcp}.
Similar approaches have also been suggested in the field
of rare-events sampling, to separate the oscillations
of the microscopic degrees of freedom from those of a few  selected slow
reaction coordinates~\cite{freeenergy}.

Controlling the temperature in these CP-like techniques
requires that one acts separately on the ionic degrees of freedom, which 
must sample the correct canonical ensemble, and on the variational
parameters, which must always remain at low temperature to minimize 
the error in the forces~\cite{bloc-parr92prb}.
Traditional stochastic thermostats allow for a highly ergodic sampling 
of all the degrees of freedom, irrespectively of their frequency. 
This is beneficial for the ionic DOFs but causes the breakdown of adiabatic 
separation. 
For this reason, deterministic thermostats of the Nos{\'e}-Hoover (NH) type~\cite{NOSE}
have been adopted. However the original NH thermostat has well-known ergodicity problems,
and the extension to Nos{\'e}-Hoover chains is normally used~\cite{CHAINS}.
This comes though at the price of introducing a large number of parameters, whose effect
on the ions dynamics is  not easy to predict and control.
 In the following we show that by using correlated noise
it is possible to tune the coupling of a stochastic thermostat
with the various degrees of freedom. This allows one not only to use
Langevin dynamics in CP-like methods, but also significantly
improves the sampling of the target ensemble, because the thermostat is tailored
to the system under study, in a predictable and controlled fashion.

We consider here a system described by coordinates $q_i$, momenta
$p_i$ and masses $m_i$, interacting via a potential $U(q)$,
where $q$ is the set of $q_i$'s.
The colored Langevin equations~\cite{zwan61pr,zwan+01book} read
\begin{equation}
\label{eq:gle}
\begin{split}
\dot{q}_i(t)&=p_i(t)/m_i \\
\dot{p}_i(t)&=f_i[q(t)]-\int_0^t\mathrm{d}t'\mathcal{K}(t-t')p_i(t')+\zeta_i(t)
\end{split}
\end{equation}
where $f_i=-\partial U /\partial q_i$ are the forces,
$\mathcal{K}(t)$ is the memory kernel and
${\zeta}(t)$ is a vector of independent Gaussian noises.
In order to set the temperature to a chosen value $T$,
the noise term ${\zeta}(t)$ needs to be related to the memory kernel by the 
fluctuation-dissipation theorem
$
\langle
\zeta_i(t)
\zeta_j(t')
\rangle
=
\delta_{ij}m_iT\mathcal{K}(t-t')
$.

The non-Markovian Eqs.~\eqref{eq:gle} might seem
at first too complex to be used in practical applications. 
However, for a rather general form of the memory kernel, 
$\mathcal{K}\left(t\right)=\Re\sum_k c_k e^{-t\left(\gamma_k+\mathrm{i}\omega_k\right)}$
with $\gamma_k>0$,
it is possible to rewrite Eq.~(\ref{eq:gle}) in an equivalent Markovian
form by introducing a set of auxiliary momenta~\cite{marc-grig83jcp,lucz05chaos}:
\begin{equation}
\begin{split}
 \dot{q}_i(t)=&s_{0i}(t)/m \\
 \dot{\mathbf{s}}_i(t)=&\left(f_i[q(t)],0,\ldots,0\right)^T- 
	\mathbf{A}\mathbf{s}_i(t)+\mathbf{B} \boldsymbol{\eta}_i(t).
\end{split}
\label{eq:generalized}
\end{equation}
Here $\mathbf{s}_i=\left(p_i,s_{i1},\ldots,s_{iN}\right)^T$ is a $N+1$ dimensional vector,
whose first component is the canonical momentum $p_i$ associated to the
$i$-th DOF, and $\boldsymbol{\eta}_i$ is a vector of Gaussian white noises,
with
$
\langle
{\eta}_{ik}(t)
{\eta}_{jk'}(t')
\rangle
=
\delta_{ij}\delta(t-t')\delta_{kk'}
$.
The real-valued matrices $\mathbf{A}$ and $\mathbf{B}$ determine the dynamics of $p_i$,
and can be related to $\mathcal{K}\left(t\right)$ by extending the arguments of Ref.~\cite{marc-grig83jcp},
as will be discussed elsewhere.

In order to illustrate some of the effects of using a
colored noise, we study the simple case in which
\begin{equation}
\!\mathbf{A}=\frac{1}{\tau_F}
\left(\!\begin{array}{cc}
       0 & -\sqrt{\gamma\tau_F} \\
	\sqrt{\gamma\tau_F} & 1
\end{array}\!\right),
\,
\mathbf{B}=
\sqrt{\frac{2 T m_i}{\tau_F} }
\left(\!\begin{array}{cc}
	0 & 0 \\
 	0 & 1
\end{array}\!\right). \label{eq:simplepars}
\end{equation}
This choice leads to the stationary distribution
\begin{equation*}
\bar{P}(q,p,s_1)\propto
\exp\!\left[
-\frac{1}{T}\!\left(
\frac{p^2}{2m}+\frac{s_1^2}{2m}+U(q)
\right)\!
\right],
\end{equation*}
corresponding to the desired canonical ensemble for $q$ and $p$.
The memory kernel and its power spectrum are
\begin{equation}
\mathcal{K}(t)=\frac{\gamma}{\tau_F}e^{-|t|/\tau_F}
\quad
\text{and}
\quad
\mathcal{S}(\omega)=\frac{\gamma}{\pi}\frac{1}{1+\tau_F^2\omega^2}
\label{eq:mem-kernel}
\end{equation}
 respectively. Thus the friction $\gamma$ determines the intensity of the kernel and
$\tau_F$ the autocorrelation time of the noise. For the purpose of this 
work, one can consider $S\left(\omega\right)$ to be a low-pass filter for the
noise, which has the cutoff frequency $\tau_F^{-1}$. 
Clearly, when $\tau_F\rightarrow 0$ the white-noise limit is recovered.

\begin{figure} [pbt]
\centering\includegraphics[clip,width=1.0\columnwidth]{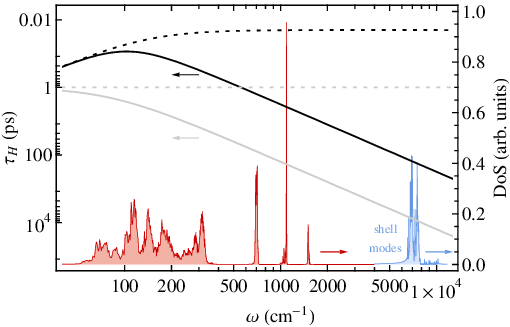} 
\caption{
\label{fig:shell-spectrum} (color online)
The autocorrelation time of the total energy for harmonic oscillators of frequency $\omega$ 
[Cf. Equation~\eqref{eq:autocorrelation-times}] is plotted
for different values of the thermostat parameters. 
Dark curves correspond to high friction ($\gamma^{-1}=20$\ fs) whereas
light ones correspond to a more gentle thermostat ($\gamma^{-1}=1$\ ps).
Dotted lines correspond to white noise ($\tau_F=0$) and full ones to 
colored noise with $\tau_F=2$\ fs.
The curves are superimposed on the vibrational density of states (DoS) for a polarizable force-field
simulation of crystalline calcite, which was obtained from the Fourier transform of the
velocity-velocity autocorrelation function. For reference, we report the shell vibrational modes
as obtained from a run where we artificially heated the shells to $300$~K.
}
\end{figure}

We consider the dynamics of a set of harmonic oscillators.
In this case Eqs.~\eqref{eq:generalized} are fully linear,
and the autocorrelation time for the total energy of an eigenmode of 
frequency $\omega$ can be explicitly evaluated~\cite{zwan+01book,gard03book}:
\begin{equation}
\label{eq:autocorrelation-times}
\tau_H\left(\omega\right)=
\frac{\gamma}{4\omega^2} + \frac{1}{\gamma}+\frac{\omega^2\tau_F^2}{\gamma}.
\end{equation}
We take $\tau_H\left(\omega\right)$ as a measure of the time needed for the thermalization
of each individual normal mode.
For a white noise ($\tau_F=0$), $\tau_H$ decreases
with $\omega$ until it reaches a plateau at $\tau_H=1/\gamma$,
while for $\tau_F\neq 0$, the autocorrelation time has a minimum at
$\omega=\sqrt{\gamma/(2\tau_F)}$ and grows quadratically
thereafter. By properly adjusting $\tau_F$, one can 
select which modes are going to be maximally coupled with the thermostat,
and thus reduce the coupling of the thermostat to the fastest modes
(see also Fig.~\ref{fig:shell-spectrum}).

We next consider the application of the colored-noise thermostat, with the 
parameters of Eq.~(\ref{eq:simplepars}), to classical MD simulation
using a polarizable force field.
Here the electronic DOF are represented by charged shells,
bound with harmonic potentials to the corresponding atomic cores.
We couple a colored-noise thermostat to the ions, at the target temperature, and
choose the filtering time $\tau_F$ in such a way that the impact on the electronic
DOF is minimal. At the same time, we apply a zero temperature thermostat of  friction $\gamma_S$ to the electrons.
This latter thermostat is memory-less, so that it couples optimally with the fast electronic modes.
Such a simulation scheme amounts to a non-equilibrium dynamics, in which heat is injected 
into the ionic DOF and systematically subtracted from the electronic ones.
In spite of the stochastic nature of these equations it is still possible to introduce a 
conserved quantity that measures the accuracy of the integration. This can be obtained
by accumulating the change in kinetic energy due to the thermostat~\cite{buss+07jcp,buss-parr07pre,brun+07jpcb,ensi+07jctc}.
However, at variance with Refs.~\cite{buss+07jcp,buss-parr07pre}, the conservation of this 
quantity does not rigorously measure the sampling accuracy.

\begin{figure} [pbt]
\centering\includegraphics[clip,width=1.0\columnwidth]{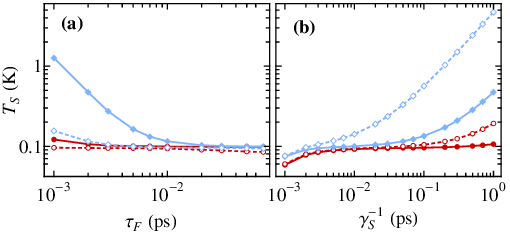} 
\caption{
\label{fig:shell-panels} (color online)
Shell temperature ($T_S$) for calcite as a function of the thermostat 
parameters. Simulations have been performed for the discrete series of values 
indicated by arrows on the horizontal axes. The points are joined by continuous lines, for clarity sake.
In both panels we distinguish the strength of the ion thermostat by
the line color. Darker or lighter (red or blue in the online version) curves
correspond respectively to a strong ($\gamma^{-1}=20$~fs) or mild ($\gamma^{-1}=1$~ps)
friction. In panel (a) we plot $T_S$ against $\tau_F$, and we choose two extreme values
of the shell friction, $\gamma_S^{-1}=1$~ps and $\gamma_S^{-1}=50$~fs, which are 
represented respectively with full and dashed lines.
In panel (b) we plot the dependence of $T_S$ versus $\gamma_S$. Here
full and dashed lines correspond respectively to a physically meaningful filter 
($\tau_F=2$~fs) and to white noise ($\tau_F=0$).
}
\end{figure}

As an example we consider the simulation of crystalline calcite, 
modeled by a polarizable force field~\cite{brun+07jpcb}.
The $\mathrm{Ca^{2+}}$ ions are treated as non-polarizable, while the polarization of the 
$\mathrm{CO_3^{2-}}$ anions is described by a charged shell attached to each oxygen.
The thermostats are applied to the non-polarizable ions and, in the case of the oxygens, to the center of mass of the 
system formed by the ion plus its shell. 
Meanwhile, the electronic temperature is controlled by the damping of the velocity 
of the shells relative to the partner $\mathrm{O}$ ions.
The vibrational density of states in the absence of any thermostat can be used
as an approximate guide to the choice of the colored thermostat parameters 
(see Fig.~\ref{fig:shell-spectrum}).
In real-life, anharmonicity will introduce some coupling between the normal modes, 
so that deviations from the predictions of Eq.~(\ref{eq:autocorrelation-times}) are expected.
However, at least in the case of quasi-harmonic modes, 
they will most likely reduce $\tau_H\left(\omega\right)$.
Thus, one can safely use the analytical estimate to tune 
the thermostat parameters beforehand, without having to perform time-consuming tests on
the real system.

We simulated~\cite{DLPOLY} a box containing 96 $\mathrm{CaCO_3}$ units,
with a timestep of $1$~fs, performing $NVT$ runs with target temperature $T=300$~K.
We performed systematic tests by varying $\tau_F$, $\gamma$ and $\gamma_S$ (Fig.~\ref{fig:shell-panels}).
The averages have been computed from $1$~ns-long runs, where we
discarded the first $100$~ps for equilibration.
Within a large range of parameters, 
the procedure performs as expected: the temperature of the 
shells remains below a few~K, and the ions equilibrate to the desired 
temperature. As $\tau_F$ is set to a value different from zero, the heat transferred
to the electronic DOF is reduced. However, some care must be taken in choosing the
friction $\gamma_S$, because the shell thermostat can induce a small drag on the ions which
results in an ionic temperature lower than desired,
if not compensated by a high thermostat strength $\gamma$.
Since $\tau_H$ does not decay fast enough to zero for $\omega>\tau_F^{-1}$,
one must choose a low cutoff frequency in order not to heat up the shells.
As a consequence, the relaxation time for high-frequency phonons increases,
making the effects of shell-induced drag more pronounced.
However, the thermostat can be systematically improved by adding more
degrees of freedom, so as to obtain a more sharply defined
filter, as we will show below.

\begin{figure}[pbt] 
\centering\includegraphics[clip,width=1.0\columnwidth]{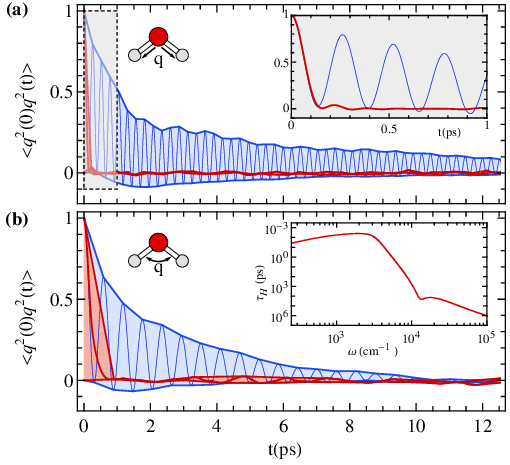} 
\caption{(color online)
Autocorrelation functions for the squares of (a) the symmetric stretching and (b) the bending modes
of a heavy water molecule in vacuum, performed in the $NVT$ ensemble at $T=300$~K. 
We use a fictitious mass $\mu=200$~a.u., and a timestep of $4$~a.u., in order to minimize the errors
on the forces~\cite{tang-scan02jcp}.
The Nos{\'e}-Hoover thermostat with chain length $4$ has been used,
and its mass chosen so as to maximize the coupling to the stretching mode.
The NH correlation functions (lighter lines, blue in the online version)
are highly oscillating and decay very slowly.
The shading highlights the curve's envelope. In contrast using the new 
thermostat (darker lines, red in the online version)
we find a much sharper decay, which in the case of the stretching 
requires an enlarged scale to be appreciated [inset of panel (a)].
In the inset of panel (b) we show the relation between $\tau_H$ and $\omega$
for our thermostat. The parameters have been optimized to obtain a sharp decay
of the response for frequencies above the stretching mode.
\label{fig:cpmd-acf}}
\end{figure}

Thermostatting on {\em ab initio} CPMD is more challenging. Since wavefunctions are not atom-centered, the 
coupling of the dynamics of the electronic DOF to the ions is stronger 
than in the shell-model case, and the presence of high-frequency components in the noise quickly
heats up the electrons. Furthermore, because of the expense of {\em ab initio} CPMD,
it is mandatory to have fast equilibration and sampling.
We will show that both problems can be solved thanks to the 
tunability and predictability of our scheme. 
As a test example, we ran simulations of a single heavy water molecule in vacuum, using a standard literature
setup (see Fig.~\ref{fig:cpmd-acf} and Ref.~\cite{sit-marz05jcp}).
We ran several independent trajectories for a total of $90$~ps, 
starting from ionic configurations equilibrated at $300$~K
and from wavefunctions quenched to the Born-Oppenheimer surface~\cite{CPMD}.
We have used Eq.~\eqref{eq:generalized} with $5$ extended momenta and fitted $\mathbf{A}$ and $\mathbf{B}$
in order to obtain a short, optimal response time over the ionic degrees of freedom, 
and an abrupt increase in the region corresponding to electronic modes 
[see inset of Fig.~\ref{fig:cpmd-acf}(b)].
We then compare this case with results from a massive Nos\'e-Hoover-chains simulation~\cite{CHAINS,tobi+93jpc}.
In both cases the strength of the thermostat is such that 
the underlying dynamics of the ions is severely altered.

With the present, very conservative choice of parameters the drift in electronic energy
is negligible for both thermostats. In Figure~\ref{fig:cpmd-acf} we 
plot the autocorrelation function of the squares of the normal modes. 
The integral of these functions measures the time
required to lose memory of the initial configuration.
It is evident that the use of an optimized colored-Langevin thermostat
dramatically reduces this time.

The thermostat we have presented offers a number of advantages.
It can be used in CP-like, extended-Lagrangian simulations, 
and it is also much faster in reaching equilibrium than the Nos{\'e}-Hoover thermostat.
This is particularly relevant when performing expensive, {\em ab initio} simulations,
but any problem which requires averaging over uncorrelated configurations
of the system can greatly benefit from the enhanced relaxation time.
The optimal parameters of the simulation can be easily estimated before the
run is started.
Here, in the difficult case of a molecule in vacuum, we have been able to reduce the 
correlation time down to a fraction of a picosecond.
An additional advantage is that the exact propagator in the case of zero force 
is obtained easily~\cite{buss-parr07pre}, which makes the implementation
simple and robust, at variance with Nos{\'e}-Hoover chains~\cite{CHAINS} which requires
a high order integrator to ensure accurate trajectories~\cite{tuck-parr94jcp}.
Finally, the introduction of highly tunable, non-Markovian thermostats in molecular dynamics 
simulations lays the foundations for the development of optimal sampling algorithms, 
which can be of great benefit in free-energy techniques, or when one must
treat systems with a broad vibrational spectrum, which is the case for instance in path-integrals MD. 
We believe that this is only a first example, and that colored noise will 
find many other applications in a variety of computational problems.
We thank Dr. G. Tribello for helping us in the simulation of
calcium carbonate and for carefully reading the paper.

\end{document}